\documentclass[aps,prb,twocolumn,superscriptaddress,amsmath,amssymb,fleqn]{revtex4}
\usepackage{graphicx}
\usepackage{mathrsfs}
\usepackage{bm}
\usepackage{verbatim}
\usepackage{dcolumn}
\usepackage{bm}
\usepackage{epsfig}
\usepackage{subfigure}

\begin{document}
\title{Current partition at topological zero-line intersections}
\author{Zhenhua Qiao}
\affiliation{Department of Physics, The University of Texas at
Austin, Austin, Texas 78712, USA}
\author{Jeil Jung}
\affiliation{Department of Physics, The University of Texas at
Austin, Austin, Texas 78712, USA}
\author{Chungwei Lin}
\affiliation{Department of Physics, The University of Texas at
Austin, Austin, Texas 78712, USA}
\author{Allan H. MacDonald}
\affiliation{Department of Physics, The University of Texas at
Austin, Austin, Texas 78712, USA}
\author{Qian Niu}
\affiliation{Department of Physics, The University of Texas at
Austin, Austin, Texas 78712, USA}
\affiliation{International Center for Quantum Materials, Peking University, Beijing 100871, China}
\date{\today}

\begin{abstract}
An intersection between one-dimensional chiral acts as a topological current splitter.
We find that the splitting of a chiral zero-line mode obeys very simple,
yet highly counterintuitive, partition laws which relate current paths to the geometry of the intersection.
Our results have far reaching implications for device proposals based
on chiral zero-line transport in the design of electron beam splitters and
interferometers, and for understanding transport properties
in systems where multiple topological domains lead to a
statistical network of chiral channels.
\end{abstract}

\maketitle
{
A massive chiral two-dimensional electron gas (C2DEG) has a valley Hall conductivity that has the same sign as its mass.
The valley Hall effect leads to conducting edge states and also, when the mass parameter varies spatially, to conducting states localized along mass zero-lines.
~\cite{highway,morpurgo,kinkstate-jeil} Provided that inter-valley scattering is weak, zero-line state properties are closely
analogous~\cite{highway,morpurgo,kinkstate-jeil} to edge state properties of quantum spin-Hall insulators
and include both chiral propagation and suppressed backscattering.\cite{highway}
Metallic zero-line modes (ZLMs), or topological 1D kink states, provide a two dimensional realization
of Dirac zero energy modes,\cite{jackiw,schrieffer}
and their existence has been proposed in a wide variety of systems including graphene mono and bilayers,
\cite{highway,morpurgo,Semenoff,YaoWang,kinkstate-jeil,killi_superlattices}
topological insulators with lattice dislocations,\cite{yingran}
boron nitride crystals with grain boundaries,\cite{nanoroad}
superfluid $^3$He, \cite{volovik}
and photonic crystals.\cite{photonic1,photonic2}
In the present Letter we examine current partition properties
at zero-lines intersections,\cite{highway,morpurgo}
which are expected to be ubiquitous in systems
in which the mass term results from a disorder potential or from
spontaneous symmetry breaking.}

\begin{figure*}
\includegraphics[width=18cm,angle=0]{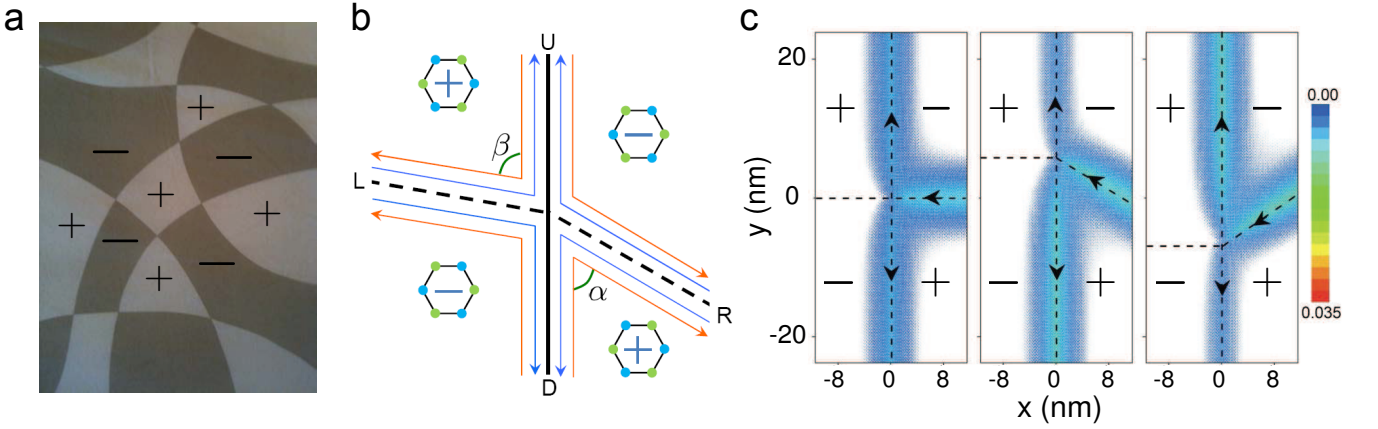}
\caption{\textbf{Current partition at zero-line intersections:}
\textbf{a}. Staggered sub lattice-potential, and hence valley Hall conductivity,
spatial distribution pattern which defines the zero-line paths.  Blue dots indicate positive
site energies and green dots negative site energies.
\textbf{b}. Schematic of four-terminal graphene samples with different staggered potential
distributions.  The Left (L), right (R), up (U), and down (D) leads are extend indefinitely from the
plotted central scattering region.
The U and D ZLMs are fixed along the vertical direction and
the angles $\alpha$ and $\beta$ specify the R and L ZLM directions.
The solid and dashed lines in black denote zero lines.
The blue and orange lines represent allowed chiral propagation paths.
\textbf{c}. ZLM LDOS distribution for modes incident from lead R
for $\alpha=90^\circ,60^\circ,120^\circ$ at fixed $\beta=90^\circ$. }\label{Figure1}
\end{figure*}

ZLMs in C2DEGs are centered on zero-lines of the mass~\cite{highway,kinkstate-jeil,morpurgo,Semenoff,YaoWang},
{\it i.e.} on lines along which the mass changes sign as illustrated in Figure \ref{Figure1}{\textbf {a}}.
A mass term leading to a valley Hall effect~\cite{xiaodi,lijian}
can be produced by a sublattice staggered external
potential in single layer graphene,~\cite{Semenoff,YaoWang}
and more practically by a gate controlled
interlayer potential difference in Bernal bilayer and ABC stacked
multilayer graphene.~\cite{highway,morpurgo,killi_superlattices,kinkstate-jeil}
Mass terms can also be generated by spin-orbit coupling~\cite{QAHEzhenhua,QSHKaneMele,tizhenhua}
and by electron-electron interactions.~\cite{fan,lattice,levitov}
In this last case ZLMs~\cite{lutliq} appear naturally at domain walls separating regions with
different local anomalous, spin, or valley Hall conductivities.

Chiral propagation implies that ZLMs can travel only in the direction which places
negative masses either on their left, or depending on valley, on their right.
It follows, as illustrated in Figure \ref{Figure1}{\textbf {c}}, that there is no forward propagation at a
zero-lines intersection; a propagating mode
is split between a portion that turns clockwise and a portion that turns counterclockwise.
These unusual transport properties are potentially valuable for new types of electronic devices.
We have therefore carried out quantum transport calculations for an explicit
model of intersecting ZLMs in order to discover rules for current partitioning
at such a ZLM splitter.  The system we study is a $\pi$-band tight-binding model
for single-layer graphene with a position-dependent sublattice-staggered potential
constructed to form intersecting zero lines which enable propagation to
four ZLMs labeled left~(L), right~(R), up~(U)
and down~(D) in Figure~\ref{Figure1}{\textbf b}.
For simplicity, we consider the case where the U and D ZLMs
are fixed along the vertical direction, and we define the angles
between R and D ZLMs to be $\alpha$
and between L and U ZLMs to be $\beta$.
The blue and orange lines in Figure \ref{Figure1}{\textbf b} indicate the allowed
chiral propagation paths.

The numerical results reported on below are for a
$\pi$-orbital tight-binding Hamiltonian with nearest neighbor hopping and
a sublattice-staggered potential:
$H =-t\sum_{\langle ij \rangle} ~ c_i^{\dag} c_j
+ U_A \sum_{i,A} c_{i}^{\dag} c_{i}
+ U_B \sum_{i,B} c_{i}^{\dag} c_{i}$.
Here $c^{\dag}_i$~($c_{i}$) is a creation~(annihilation)
operator for an electron at site $i$,
and $t = 2.6$ eV is the nearest neighbor hopping amplitude.
For a sublattice staggered potential the A and B sublattice energies are
opposite, {\it i.e.} $U_A=-U_B=\lambda U_0$, where $U_0$ measures the potential
strength and $\lambda$=``$\pm$"
determines the sign of the valley Hall effect in each quadrant.
In all our simulations, the potential amplitude was chosen to
be $U_0/t=0.05$.
The ZLMs appear confined along zero-lines where the mass
becomes zero and the resulting wave-function tails spread into the bulk
with a depth proportional to the inverse of the mass \cite{highway, morpurgo}.

Although our study has been carried out in single layer graphene for computational convenience,
we expect that similar conclusions apply to ZLMs in other systems, and in particular in
bilayer graphene where the spatially patterned mass term can in principle be generated
externally using gates.
Our transport calculations are based on the
Landauer-B\"{u}ttiker formalism \cite{datta} and recursively constructed
Green functions.\cite{transfer1,transfer2}
The conductance from lead $q$ to lead $p$ is
numerically evaluated from
${{\rm G}_{pq}}=({e^2}/{h}){{\rm Tr} [\Gamma_{p} {\rm G}^r \Gamma_{q} {\rm G}^a]}$,
where $e$ is the electron charge,
$h$ is the Planck's constant, ${\rm G}^{r,a}$ are the retarded and
advanced Green's functions,~\cite{datta} and ${\Gamma}_{p}$ is a linewidth function describing the
coupling between lead $p$ and the central region.
The propagation of ZLMs, or kink states, incoming from lead $p$ is effectively illustrated by
plotting a map of its contribution to the local density of states (LDOS) at an energy $\varepsilon$
in the gap: $\rho_p({\bf r},\varepsilon)={1}/{2\pi}[{{\rm G} ^r \Gamma _p {\rm G}^a]_{\bf r r}}$.
Here ${\bf r}$ is the real space coordinate.

The central scattering region in our calculations is rectangular with
size $n_x=94$ and $n_y=432$ as explained fully in the Supplementary Information.
The valley label of a state is of course not a good quantum number for ZLMs and
valleys are most strongly mixed when their wave vector projections in the propagation
direction are identical.  For energies inside the gap this coincidence happens only for
propagation in the armchair
direction\cite{Semenoff,highway,kinkstate-jeil,nanoroad}.
However, numerical calculations have shown
a remarkable absence of back-scattering at sharp turns in the zero-line or
at the encounter of a ZLM splitter \cite{highway}
except in a narrow energy range very close to the avoided crossing gap centered on $\varepsilon/t=0.00$
between modes with opposite propagation directions.
For the results shown below we have chosen $\varepsilon/t=0.01$
to avoid this energy range; the chirality of the ZLM modes is then
very well defined.

In a four terminal ZLM splitter device [see Figures~\ref{Figure1}{\textbf b}],
there are in total twelve distinct inter-terminal conductance values.
The number of independent conductances is reduced to six
in time-reversal symmetric systems
since ${\rm G}_{pq}={\rm G}_{qp}$.
For chiral transport forward scattering and back scattering are absent at
a ZLM intersection, reducing the number of
independent parameters further.  The current conservation then implies that
${\rm G}_{pr}+{\rm G}_{qr}={\rm G}_0= e^2/h$
for any value of $r$ and $p,q$ the labels of the two neighboring leads.
It follows that
\begin{eqnarray}
{\rm G_{LU}} &=& {\rm G_{RD}} \quad \& \quad
{\rm G_{RU}} =  {\rm G_{LD}},   \label{relation1}
\end{eqnarray}
and that ${\rm G_{RU}} + {\rm G_{RD}}= e^2/h$,
leaving only one independent parameter for the entire four terminal systems.
In a ZLM splitter with zero backscattering and perfect chiral current filtering
transport is completely characterized by specifying how incoming current at an intersection
is partitioned between clockwise and counterclockwise rotation outgoing directions.
The partition law must be the same for all incoming channels.
In the following, we focus on the conductances $\rm G_{UR}$ and $\rm G_{DR}$
corresponding to the currents incoming from lead R.
The above relations were numerically verified for a ZLM
current splitter with $\alpha = \beta = 90^{\circ}$
in Ref.~[\onlinecite{highway}] and we have now numerically verified that they are true
for arbitrary values of the lead angles $\alpha$ and $\beta$.
(See the Supplementary Information for further details.)

\begin{figure}
\includegraphics[width=8.5cm,angle=0]{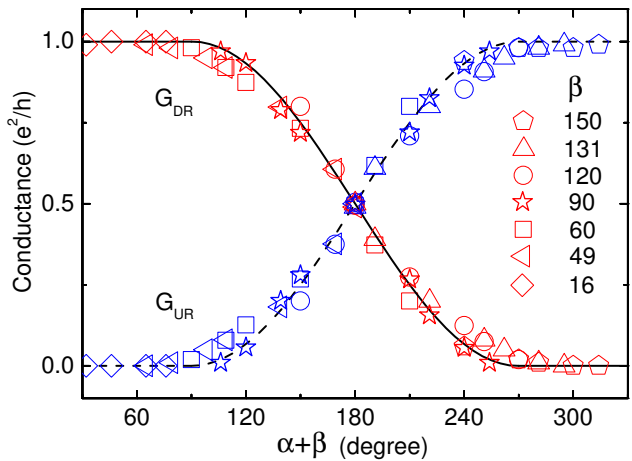}
\caption{
\textbf{Current partition law:} Current as a function of angles $\alpha$ and $\beta$
for the zero-line intersection defined in Fig. 1c.
The current partition follows a simple relation given by Eq. \ref{partition}
whenever $90^{\circ} \leq \alpha + \beta \leq 270^{\circ}$
and completely follows the sharper rotation path
outside this range.
The solid and dashed lines are the fitting functions
defined in Equation~(\ref{partition}).
}
\label{Figure2}
\end{figure}

For a ZLM splitter with a fixed vertical pair of leads U and D and rotatable R and L leads as shown in Figure \ref{Figure1}{\textbf b}, we have numerically discovered a rather simple law which describes the dependence of the
current partition on the angles $\alpha$ and $\beta$ with surprising accuracy:
(See Figure~\ref{Figure2} for a summary of the data which supports this law and
the Supplementary Information for further details.)
\begin{eqnarray}
{\rm G_{UR}}&=&\frac{{\rm G_0}}{2} \left[ 1 - \sin \left( \alpha + \beta \right) \right], \nonumber \\
{\rm G_{DR}}&=&\frac{{\rm G_0}}{2} \left[ 1 + \sin \left( \alpha + \beta \right) \right],  \label{partition}
\end{eqnarray}
for $90^{\circ} \leq \alpha + \beta \leq 270^{\circ}$.
Outside of this angle range the current follows the path with the larger rotation angle:
${\rm G_{DR}} = {\rm G_{UL}} = {\rm G_0}$ and
${\rm G_{UR}} = {\rm G_{DL}} = 0$ when $0^{\circ} < \alpha + \beta < 90^{\circ}$,
or
${\rm G_{UR}} = {\rm G_{DL}} = {\rm G_0}$ and
${\rm G_{DR}} = {\rm G_{UL}} = 0$ when $270^{\circ} < \alpha + \beta <  360^{\circ}$.
For the special case of $\beta = 90^{\circ}$ the current partition law simplifies to
${\rm G_{UR}} = {\rm G_0} \sin^2 (\alpha/2)$ and
${\rm G_{DR}} = {\rm G_0} \cos^2 (\alpha/2)$.
And for the special case of $\alpha = \beta$, corresponding to zero lines that are
straight at the intersection point and therefore to a mass pattern defined by a
smooth external potential,
${\rm G_{UR}} = {\rm G_0} \sin^2 (\alpha-45^\circ)$ and
${\rm G_{DR}} = {\rm G_0} \cos^2 (\alpha-45^\circ)$.

Contrary to intuition, as illustrated in Figure \ref{Figure1}{\textbf c} for a current incoming from lead R,
the current partition law implies that more current always follows
the path requiring a larger rotation angle.
This makes sense since this pair of incoming and outgoing modes interact
over a longer distance, providing more opportunity for interactions which
lead to inter-mode scattering (The zero-line mode decay length
is $\sim m v_{D}/\hbar$ where $m$ is the local mass and $v_{D}$ is the
Dirac point velocity for $m \equiv 0$.~\cite{Semenoff}).
It is noteworthy that the current partition depends only on the combination
$\alpha + \beta$ and indeed it is surprising that the angle $\beta$, which specifies
the orientation of a lead that does not
carry any current, influences the partition law.
This law evidently results, however, from the interference between ZLMs
close to the ZLM intersection point, allowing inactive leads to play a role in the
scattering. Note that the completely unbalanced current partition favored by a
very small value of $\alpha$ (a sharp turn from R to D) is mitigated
if $\beta$ (the L to U turn angle) is larger than $90^\circ$.

We emphasize we have not derived the form which fits our current partition numerical results
analytically and it is likely only approximate.   It applies accurately only
at energies close to the middle of the bulk energy gaps.
The expression can be rationalized by the following argument.
Let us consider a ZLM splitter, where the chiral current filtering rule restricts
current incoming from R or L to scatter
into the U and D leads.
By denoting the asymptotic amplitudes of the ZLM
at the leads as $A_R$, $A_L$, $A_U$, and $A_D$,
current conservation implies that
\begin{eqnarray}
\label{scattering}
\left[
\begin{array}{l}
 A_{U} \\
 A_{D}
\end{array}
\right]=\left[
\begin{array}{ll}
 \cos(\tau)e^{iu} & -\sin(\tau)e^{-iv} \\
 \sin(\tau)e^{iv} & \cos(\tau)e^{-iu}
\end{array}
\right] \left[
\begin{array}{l}
 A_{L} \\
 A_{R}
\end{array}
\right].\label{AnalyticEquation}
\end{eqnarray}
Equation~(\ref{AnalyticEquation}) relates the amplitudes of incoming and outgoing waves via
a general SU(2) unitary transformation matrix with parameters $u$, $v$, and $\tau$.
When we set $A_L$ to be zero,
then for any $A_R$, $|A_U|^2=|A_R|^2 \sin^2 \tau$, and $|A_D|^2=|A_R|^2 \cos^2 \tau$.
The scattering phases $u$ and $v$ are irrelevant for the scattering probability
and will henceforth be dropped.
The conductances from R can thus be written as
${\rm G_{UR}}={\rm G_0} \sin^2 \tau(\alpha,\beta)$
and ${\rm G_{DR}}={\rm G_0} \cos^2 \tau(\alpha,\beta)$,
where ${\rm G_0}=e^2/h$.
Equipartition for $(\alpha,\beta)=(\alpha,180^\circ - \alpha)$ implies that
$\tau(\alpha,180^\circ - \alpha)=45^\circ$.
For the current coming from R into D lead,
we get from symmetry considerations the relation
${\rm G_{DR}}(\alpha,\beta) = {\rm G_{DR}}(\beta,\alpha)$.
If we assume that $\tau$ is a smooth function of
$\alpha$ and $\beta$ and additionally require that the conductances are invariant
under $(\alpha,\beta)  \to (\alpha + N\times360^\circ,\beta + M\times 360^\circ)$
we can conclude that $\tau(\alpha,\beta) =  c(\alpha + \beta)/2 - 45^\circ$, where $c=\pm 1$. Then we impose a second condition for the current partition saturation when $\alpha+\beta=90^\circ$, for instance in the limit when $\beta=90^\circ$ and $\alpha=0^\circ$ where R and D ZLMs merge together,
that further restricts $c=+1$. This expression reproduces
Equation~(\ref{partition}) for ${\rm G_{DR}}$ and ${\rm G_{UR}}$,
outside the saturation region which occurs for $\alpha+\beta < 90^\circ$ and $\alpha + \beta > 270^\circ$.

\begin{figure}
\includegraphics[width=6cm,angle=0]{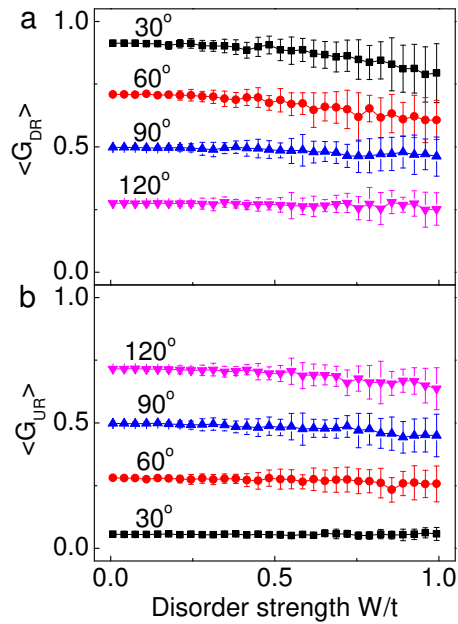}
\caption{\textbf{Influence of disorder on the current partition law:}
Averaged conductances $\langle \rm G_{DR} \rangle$ (in panel \textbf{a})
and $\langle \rm G_{UR} \rangle$ (in panel \textbf{b}) from lead R to leads
D and U as a function of disorder strength $W$ at fixed $\beta=90^\circ$.
Four representative angles of incidence $\alpha=30^\circ,~60^\circ,~90^\circ$,
and $120^\circ$ are considered.
Over 50 samples are collected at each point.
The error bar is used to indicate the strength of fluctuations as a
function of disorder realization.} \label{Figure3}
\end{figure}

Now we examine the robustness of our results in the presence of disorder.
Long-range disorder is not effective in producing inter valley scattering\cite{highway}
so we focus on the potentially more important
short-range disorder which we model as a
random potential at each lattice site.
Specifically we add a term $H_{dis}=\sum_i {\omega_i c^\dag_i c_i}$ with $\omega_i$ being
uniformly distributed in the interval of $[-W/2,+W/2]$, where $W$ characterizes the disorder strength.
In our simulations we considered over 50 realizations of the random disorder potential for
each value of the strength.  Figure~\ref{Figure3}
illustrates our results for the average conductances
$\langle {\rm G_{UR}} \rangle$ and $\langle {\rm G_{DR}} \rangle$ as a function of disorder strength $W$.
We see that the current partition law remains accurate
up to disorder strengths larger than the bulk band gap $\Delta/t=0.1$.
When the disorder strength is further increased,
the averaged conductances $\langle \rm G_{ UR} \rangle$ and $\langle \rm G_{ DR} \rangle$
are slightly reduced and disorder fluctuations grow.
For example, when the disorder strength reaches $W/t=1.0$,
10 times larger than the bulk band gap, the averaged conductances
still retain over 80\% of their original values.  The lost current takes
advantage of the intervalley scattering to access the outgoing modes of the
$L$ lead or to backscatter in the $R$ lead.
All these findings strongly indicate that the current partition law is very
robust to disorder, suggesting that ideal zero-line transport
properties can be approximated in real devices.

Since there are presently no practical techniques for imposing
staggered sublattice potentials in single layer graphene, other closely related
systems may ultimately be of greater experimental interest.
In Bernal stacked bilayer graphene, for example, a ZLM splitter can be realized
by using gates to achieve perpendicular electric fields which vary in sign spatially.
Another possibility is 2D honeycomb photonic crystals, in which the Dirac
points have been experimentally observed and sublattice staggered potentials can be
realized by choosing different diameters for the cylinders which form the structure
or by varying the dielectric material used.

In summary, when intervalley scattering can be neglected, transport along the zero lines of a
sublattice-staggered potential in graphene is chiral, requiring travel in a direction which keeps
positive masses on either the left or the right, depending on valley.
We have used the Landauer-Buttiker formula and recursively constructed Green's functions
to examine how chiral currents are partitioned between available outgoing leads
at a ZLM intersection.  We find that at energies near the middle of the bulk gap
our numerical results for the dependence of current on ZLM geometry are accurately described by
a simple partition law specified in Eq.~(\ref{partition}), and that the influence of
disorder on this law is weak.
The helicity of ZLM provides a new mechanism for
allowing or blocking currents and
may find applications in alternative designs
for nanoelectronic devices or in
enabling electron quantum interferometry~\cite{Heiblum} in a new setting.
We have explored,
for the first time to our knowledge, the geometry-dependent current partition laws
at the intersection of two zero-lines.
It will be interesting to extend our present studies to more general
parameter spaces and to look for
similarities and differences with respect to other systems with
chiral 1D transport channels including photonic crystals,
quantum anomalous Hall  and quantum Hall effect systems,
and chiral superconductors.

\textbf{Acknowledgements.---}
We are grateful to Wang-Kong Tse, Wang Yao, Jian Li and Yue Yu for useful discussions
and P. J. Coulchinsky for his help with figures.
This work was financially supported by the Welch Foundation (F-1255 and TBF1473), and by
DOE (DE-FG03-02ER45958, Division of Materials Science and Engineering) grant.
Additional support was provided by grants NBRPC (2012CB-921300) and NSFC (91121004).
The Computer Center of The University of Hong Kong is gratefully acknowledged for high-performance
computing assistance [supported in part by a Hong Kong UGC Special Equipment Grant (SEG HKU09)].

\end{document}